\documentclass{aa}
\usepackage{latexsym}
\usepackage{natbib}
\usepackage{graphicx}
\usepackage{times}  \DeclareSymbolFont{operators}{OT1}{ptm}{m}{n}
\newcommand{\sect}[1]{Sect.\,\ref{#1}}

\newcommand{\fig}[1]{Fig.\,\ref{#1}}
\newcommand{\figs}[1]{Figs.\,\ref{#1}}

\newcommand{\FWHM}{{\sc{fwhm}}}

\graphicspath{{./}{figs_pdf/}}

\begin{document}

%
\title{Constant cross section of loops in the solar corona}


\authorrunning{H. Peter \& S. Bingert}

\author{H.~Peter, S.~Bingert}

\institute{Max-Planck-Institut f{\"u}r Sonnensystemforschung, 
           37191 Katlenburg-Lindau, Germany, email: peter@mps.mpg.de
           }

\date{Received 24 April 2012 / Accepted 3 August 2012}

\abstract%
%
{%
The corona of the Sun is dominated by emission from loop-like structures. When observed in X-ray or extreme ultraviolet emission, these million K hot coronal loops show a more or less constant cross section.
}
{%
In this study we show how the interplay of heating, radiative cooling, and heat conduction in an expanding magnetic structure can explain the observed constant cross section.
}
{%
We employ a three-dimensional magnetohydrodynamics (3D MHD) model of the corona. The heating of the coronal plasma is the result of braiding of the magnetic field lines through footpoint motions and subsequent dissipation of the induced currents. From the model we synthesize the coronal emission, which is directly comparable to observations from, e.g., the Atmospheric Imaging Assembly on the Solar Dynamics Observatory (AIA/SDO).
}
{%
We find that the synthesized observation of a coronal loop seen in the 3D data cube does match actually observed loops in count rate and that the cross section is roughly constant, as observed. The magnetic field in the loop is expanding and the plasma density is concentrated in this expanding loop; however, the temperature is not constant perpendicular to the plasma loop. The higher temperature in the upper outer parts of the loop is so high that this part of the loop is outside the contribution function of the respective emission line(s). In effect, the upper part of the plasma loop is not bright and thus the loop actually seen in coronal emission appears to have a constant width.}
{%
From this we can conclude that the underlying field-line-braiding heating mechanism provides the proper spatial and temporal distribution of the energy input into the corona --- at least on the observable scales.
}
%

\maketitle

\section{Introduction\label{S:intro}}

The corona of the Sun is dominated by loop structures containing plasma at temperatures of $10^6$\,K and above. These coronal loops were first revealed by coronagraphic observations in the 1940s \citep{Bray+al:1991}. Their highly dynamic nature became very clear through extreme ultraviolet (EUV) imaging by the Atmospheric Imaging Assembly \cite[AIA;][]{Lemen+al:2011} on the recent Solar Dynamics Observatory \citep[SDO;][]{Pesnell+al:2011}. The EUV and X-ray emission from the Sun is closely related to the magnetic field in the corona \citep{Poletto+al:1975} and most prominently seen above active regions hosting sunspots on the solar surface. 
Stereoscopic EUV observations, together with extrapolations of the coronal magnetic field from surface observations, have demonstrated that loops roughly follow the magnetic field \citep{Feng+al:2007}. The formation and appearance of these loops in the complex magnetic environment of the corona provides a pivotal test for a model of the coronal heating process

In the corona the energy density of the magnetic field exceeds the internal energy density of the plasma, and therefore the magnetic field shapes the structures we see in EUV and X-ray observations. In contrast to a neutral gas, in an ionized plasma the heat conduction is no longer isotropic, but basically parallel to the magnetic field. In consequence, neighboring field lines are thermally isolated. In addition, the plasma flow has to be parallel to the magnetic field, so that one can consider each field line as an independent unit. If a part of the corona is now filled with hot plasma, either by chromospheric evaporation \citep{Klimchuk:2006}, injection of heated material from below \citep{dePontieu+al:2011}, or any other mechanism, only the filled field line will be observed as a bright coronal loop. In other parts there is still a magnetic field, but not enough heating or mass injection to load enough plasma onto the field lines.

Observations show that most of these coronal loops have a roughly constant cross section or else expand only little with height. This is obvious from the inspection of AIA images in the EUV as was hinted at from first subarcesec resolution EUV and X-ray images \citep{Dere:1982,Golub+al:1990}, and confirmed quantitatively later \citep{Klimchuk:2000,Watko+Klimchuk:2000}. The constant cross section is not simply an artifact of image processing, e.g. background subtraction \citep{LopezFuentes+al:2008}.

This constant loop width is in stark contrast to the anticipated magnetic field structure. Extrapolations of the magnetic field from the solar surface into the corona show a strong expansion of the magnetic field. Just as for a magnetic dipole, in a solar active region the magnetic field lines should strongly diverge on their way into the corona. A quantitative comparison shows that the magnetic field lines derived from the extrapolation expand by a factor of two or more, while the cospatial EUV loop keeps a roughly constant cross section \citep{LopezFuentes+al:2006}. If the EUV and X-ray loops follow the field lines as outlined above, they should participate in the magnetic expansion, which they do not.

To reconcile the magnetic structure with the EUV and X-ray loops numerous suggestions have been made about how special magnetic configurations could only show  a little expansion. In force-free, highly sheered magnetic field configurations, strong field-aligned currents could prevent the magnetic expansion \citep{McClymont+Mikic:1994}. However, the required strong currents seem to contradict observations \citep{Metcalf+al:1994}. It is also unlikely that additional magnetic twist or a variable strong cross-field elongation of the magnetic flux tube defining the loop could account for its constant width \citep{Klimchuk:2000}. Likewise, for force-free twisted magnetic flux tubes with an inflow of plasma one would not expect a constant cross-section loop under solar conditions \citep{Petrie:2008}. Magnetic separators connecting magnetic null points can be expected to show less expansion with height and thus could host nearly-constant width coronal loops \citep{Plowman+al:2009}; however, it remains to be seen whether coronal loops are always associated with separators or with their generalization the (quasi-)separatrix layers within regions of strong magnetic field \citep{Priest+Demoulin:1995}. The latter ones are associated with expanding fans at the periphery of active regions \citep{Schrijver+al:2010} but have not yet been clearly related to loops.

A different approach is to use an expanding fixed magnetic field structure and solve the hydrodynamic problem with a parameterized heating function along the field, thereby providing the thermal properties to infer the emission as observed by EUV and X-ray imaging \citep{Mok+al:2005}. By optimizing the parametrization of the heating function, this procedure can give loops of roughly constant width through the convolution of the resulting temperature and density profiles \citep{Mok+al:2008}, while it seems that some fine-tuning is required to obtain this result. Therefore the logical next step is to account for the full interaction of the plasma and the magnetic field and to include a more self-consistent treatment of the energy input to sustain the hot corona.

\section{3D MHD coronal model\label{S:model}}

In the model we present in this study we incorporate a self-consistent time-dependent heating of the plasma in an evolving magnetic environment. Horizontal granular motions in the photosphere braid the magnetic field lines which leads to currents in the upper atmosphere that subsequently heat the plasma through Ohmic dissipation. This concept follows the nanoflare model \citep{Parker:1972,Parker:1983} and is similar to the flux-tube tectonics model  \citep{Priest+al:2002}. Numerical experiments using three-dimensional magnetohydrodynamics (MHD) models show that this process leads to a loop-dominated corona  \citep{Gudiksen+Nordlund:2002,Gudiksen+Nordlund:2005a,Gudiksen+Nordlund:2005b} and that many derived observable quantities match observed average properties \citep{Peter+al:2004,Peter+al:2006,Hansteen+al:2010,Zacharias+al:2011.doppler}, including the transition region Doppler shifts or the differential emission measure.

The details of our numerical model have been described before  \citep{Bingert+Peter:2011}. Basically, we solve the time-dependent 3D MHD problem, i.e.\ the induction equation along with the mass conservation, momentum, and energy balance from the cool solar surface into the hot corona above an active region. The energy equation includes Spitzer heat conduction along the magnetic field, optically thin radiative losses and Ohmic heating. Only this ensures a proper description of the coronal pressure, which is prerequisite to synthesizing EUV emission as would be expected for, say, AIA observations. For numerical stability also a (small) heat transfer perpendicular to the magnetic field is needed. In the coronal part of the computational domain the perpendicular heat conduction is about 100 or more times weaker than the parallel conduction, which ensures that neighboring magnetic structures are still thermally isolated. To solve the problem numerically we use the Pencil code\footnote{http://code.google.com/p/pencil-code/} \citep{Brandenburg+Dobler:2002}.

The computational domain covers 50$\times$50\,Mm$^2$ in the horizontal and 30\,Mm in the vertical direction. The grid spacing is 230\,km in the vertical and 390 km in the horizontal direction. For the initial condition we use the same magnetogram as \cite{Bingert+Peter:2011} and fill the volume with a magnetic field based on a potential field extrapolation. For the initial temperature we assume a simple solar-like average stratification and calculate the initial density from hydrostatic equilibrium. We then force the lower boundary by applying a granulation-type horizontal motion as described by \cite{Bingert+Peter:2011}. This has the same statistical properties as real granulation. At the bottom the magnetic field is frozen to the plasma motions, which leads to the braiding of the field lines. In the horizontal direction we employ periodic boundary conditions.

As in other models of the same type \citep[e.g.][]{Gudiksen+Nordlund:2002,Hansteen+al:2010,Bingert+Peter:2011}, Ohmic heating dissipates the induced currents $j$ and heats the corona. This is through a term $\eta\,\mu_0\,j^2$ in the energy equation with the magnetic permeability $\mu_0$ and a constant magnetic resistivity $\eta{=}10^{10}$\,m$^2$/s. As in the aforementioned models we use a value for $\eta$ so that the magnetic Reynolds number when using the grid spacing for the length scale is on the order of unity, which requires $\eta$ to be much larger than what follows from classical transport theory \citep[e.g.][]{Boyd+Sanderson:2003}. This ensures that current sheets have a thickness of at least the grid spacing, so that the numerical scheme can resolve them. On the real Sun, for much lower values of $\eta$, much thinner current sheets would form. Still, as long as the Poynting flux into the corona is the same,  the dissipated energy should remain the same, because the currents are then higher \citep{Hendrix+al:1996,Galsgaard+Nordlund:1996}. Of course when going to smaller and smaller length scales, other physical processes come into play, especially when going below the electron mean free path (several 10 km) or the electron gyro radius (several m). In this study we are interested in the observable consequences on currently resolvable scales, so one should consider the Ohmic heating term $\eta\,\mu_0\,j^2$ as a parameterization of the true heating mechanism.

\begin{figure}
\includegraphics{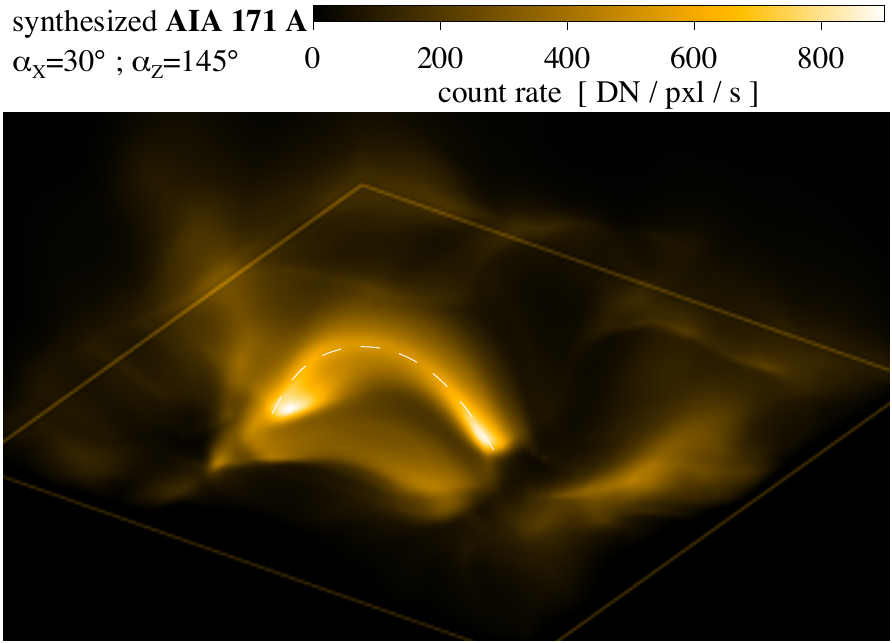}
\caption{View of the modeled corona above an active region as would be seen by AIA/SDO in the 171\,\AA\ channel dominated by emission from below about $10^6$\,K. The rectangle indicates the boundaries of the computational domain (50$\times$50$\times$30\,Mm) at the coronal base. The most prominent loop connects the periphery of the two magnetic concentrations at the surface (not shown here) and has a roughly semicircular shape with about 30\,Mm length. In the vicinity of the active region, hazy emission can be seen that is associated with the diffuse background corona in the quiet Sun.
The white dashed line indicates the position of a loop investigated  further in \fig{F:loop.var}.
The temporal evolution over 50 min is shown in a movie available in the online edition and at http://www.mps.mpg.de/data/outgoing/peter/papers/2012-loops/aa19473-peter-fig1.mp4.
\label{F:overview}}
\end{figure}

\section{Synthesized and real EUV images\label{S:loop}}

In \fig{F:overview} we display the synthesized emission in the AIA 171\,\AA\ channel from the computational domain showing plasma at just below $10^6$\,K. This image was synthesized using the AIA temperature response functions \citep{Boerner+al:2011,Peter+al:2012.loop}. 

A coronal loop sticks out, connecting the periphery of the two main polarities (sunspots) in the photosphere. This shows only a snapshot from the numerical experiment, and the structures quickly evolve in time --- some 30 minutes later other structures close-by will dominate. The image also changes with the viewing angle, of course, as a result of the line-of-sight integration of the optically thin emission. The figures shown in this paper show snapshots from the simulation, all about one hour solar time from the start of the simulation. The temporal evolution for some 50 minutes after this snapshot is  available in the online-edition.

\begin{figure}
\includegraphics{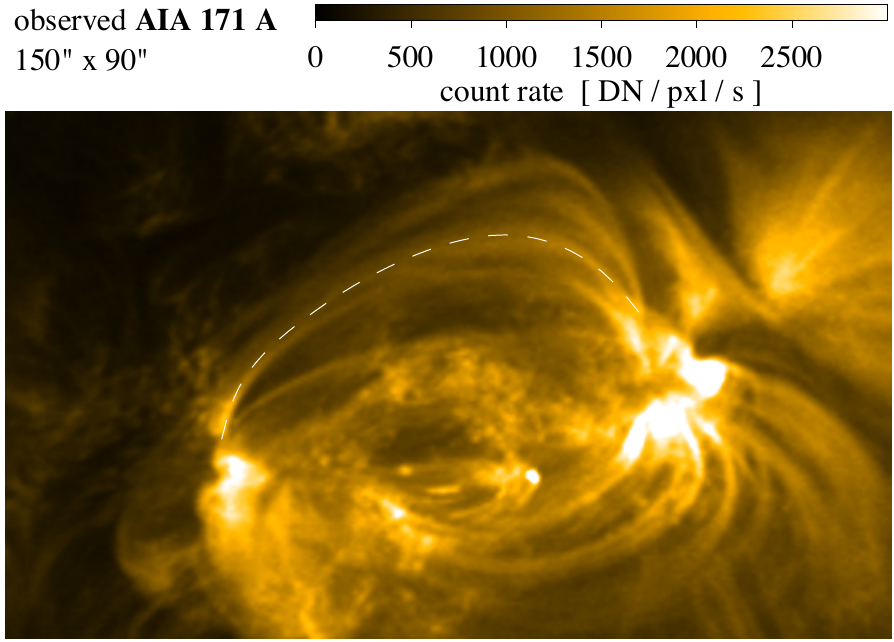}
\caption{%
Real observation of an active region.
This snapshot shows an active region observed with AIA in the 171\,\AA\ bandpass on Aug 08 2010 at 01:13:49 UTC.
The field-of-view covers 150\arcsec\,$\times$\,90\arcsec\ centered at solar (X,Y) $\approx$ (185\arcsec,\,310\arcsec). South is top.
The white dashed line indicates the position of a loop investigated further in \fig{F:loop.var}.
The temporal evolution over 75 min (from 00:36 to 01:51) is shown in a movie available in the on-line edition and at http://www.mps.mpg.de/data/outgoing/peter/papers/2012-loops/aa19473-peter-fig2.mp4.
\label{F:aia.obs}}
\end{figure}

For comparison in \fig{F:aia.obs} we show an actual observation of an active region using AIA. A system of loops connecting the main polarities of the active region is dominating the field-of-view. The numerical experiment presented in this manuscript was \emph{not} set up to match the specific observation shown here. A movie showing the temporal evolution is available in the online edition.%

\begin{figure}
\centerline{\includegraphics[width=0.90\columnwidth]{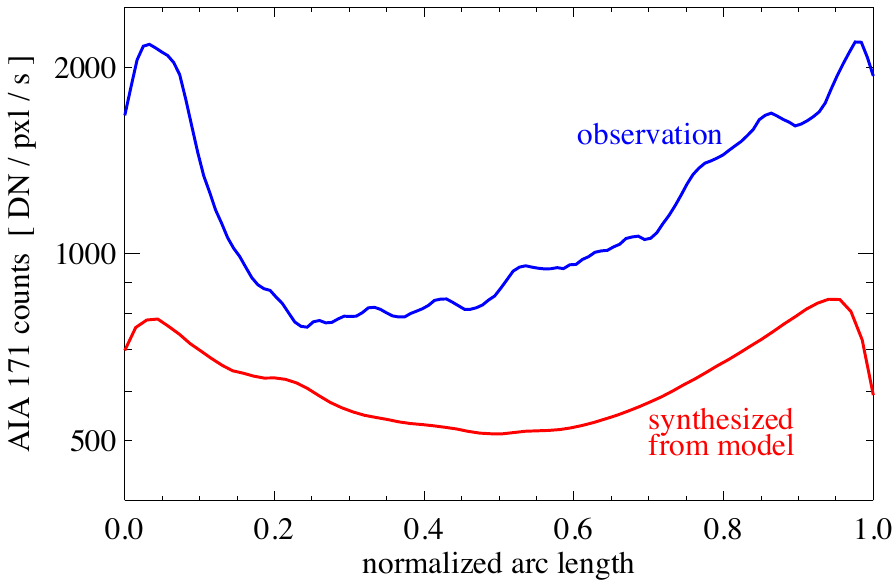}}
\caption{%
Intensity variation \emph{along} selected coronal loops. The bottom curve shows the variation along the loop synthesized from the 3D MHD model outlined in \fig{F:overview}. The top curve shows the same for the observed loop outlined in \fig{F:aia.obs}. The ordinate is plotted logarithmically.
\label{F:loop.var}}
\end{figure}

The count rates of the coronal images synthesized from the 3D MHD active region model roughly match the typical count rates actually observed in active regions with AIA. The count rates in the synthesized loop are typically a factor of two lower than in the loops of the real active region (see color scales in \figs{F:overview} and \ref{F:aia.obs} and the profile in \fig{F:loop.var}). This difference in count rate relates to a difference in density of about a factor of 1.4. This is small considering the scatter of active region brightness and the fact that we did not specifically designed the model to reproduce this particular active region. This shows that the model loop has roughly the same amount of emitting material as the real Sun, therefore the thermal structure (density and temperature) found in our model should be similar to the Sun. Besides the well defined loop we also see other structures, some with higher contrast, some appearing quite diffuse, similar to what is found in real observations. Of course, the actual active region shows a finer structure, which reflects the limitations of the current MHD models.

The temporal evolution of the loop in the synthesized corona and the real observed loops (see online material with \figs{F:overview} and \ref{F:aia.obs}) is similar in the sense that in both cases the loops have a limited lifetime of some 20 to 30 minutes. However, the real observations show a higher variability, especially on smaller scales that are not described by our model. This again reflects the limitations of the model. But still, the model and the observations have the common property of a lifetime of a larger structure of about one half hour, which is consistent with the coronal (radiative) cooling time. Further detailed work is needed, especially for more complex and longer numerical experiments, to draw further conclusions on how well the models reproduce the variability seen in coronal observations.

The intensity distribution along the loop synthesized from the model (and highlighted in \fig{F:overview}) is shown in \fig{F:loop.var}. The footpoints are slightly brighter compared to the loop apex. This can be compared to the intensity variation for one particular loop indicated in \fig{F:aia.obs}, which is overplotted in \fig{F:loop.var}.
The ratio of the count rate for the observed and synthesized loop is about a factor of two. Thus the variation along the synthesized loop roughly matches observations; however, as the loops in the model and the observation change in time, this can only be a snapshot.
\cite{Mok+al:2008} report in their loop study that ``an important deficiency of [their] model, as well as other models \citep{Schrijver+al:2004,Brooks+Warren:2006,Warren+Winebarger:2006}, is the excessive EUV emission from the footpoints.''
In the model presented in this study we have not found these overly bright footpoints, but the ratio of loop apex to footpoint is roughly comparable to observations. This is not only true for the snapshot shown in \figs{F:overview} and \ref{F:loop.var}, but it also holds for other times (see animation attached to \fig{F:overview} in the online edition).
One reason for this difference between the model of \cite{Mok+al:2008} and our model could be that theirs has a stronger concentration of the heating rate (per particle) towards the loop footpoints, which could lead to an increased radiative output near the loop feet  (see discussion in Appendix\,\ref{S:app}).

\begin{figure*}
\sidecaption
\includegraphics[width=12cm]{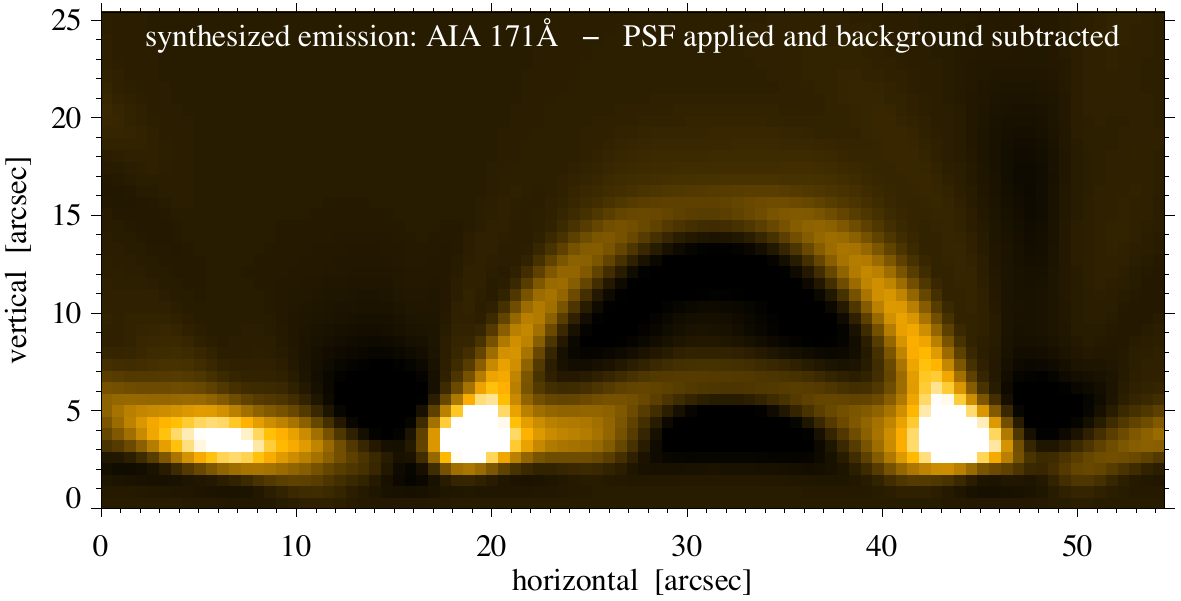}
\caption{View of the synthesized coronal loop in AIA 171\,\AA\ seen in \fig{F:overview} when horizontally integrating through the box along a line-of-sight perpendicular to the loop plane. The AIA 171\,\AA\ point spread function (PSF) was applied and shows the image with the same pixel size as AIA (0.6\,arcsec corresponding to 435 km on the Sun). As in observations, a background subtraction was applied to enhance the contrast.
\label{F:synt.aia}}
\end{figure*}

\section{Loop cross section\label{S:width}}

Already in \fig{F:overview} the loop looks like is has a more or less constant cross section. This becomes more evident when integrating through the box horizontally (along a line-of-sight perpendicular to the loop plane). This is shown in \fig{F:synt.aia} where we also applied the point-spread-function (PSF) for the AIA 171\,\AA\ channel%
\footnote{The point spread function (PSF) for AIA is available in \emph{SolarSoft} (http://www.lmsal.com/solarsoft/).}
for a better comparison with real AIA images. The image is displayed with the actual AIA pixel size and after background subtraction (as also used in observations to enhance the contrast). From this image one would roughly estimate the loop width by eye to be some four to five AIA pixels wide, corresponding to some two to three Mm. For a roughly 30\,Mm long loop (such as the one in \fig{F:synt.aia}), this is consistent with observations and with an order-of-magnitude estimation for the loop width based on the field-line-braiding mechanism \citep{Schrijver:2007}.

\begin{figure}
\centerline{\includegraphics[width=0.9\columnwidth]{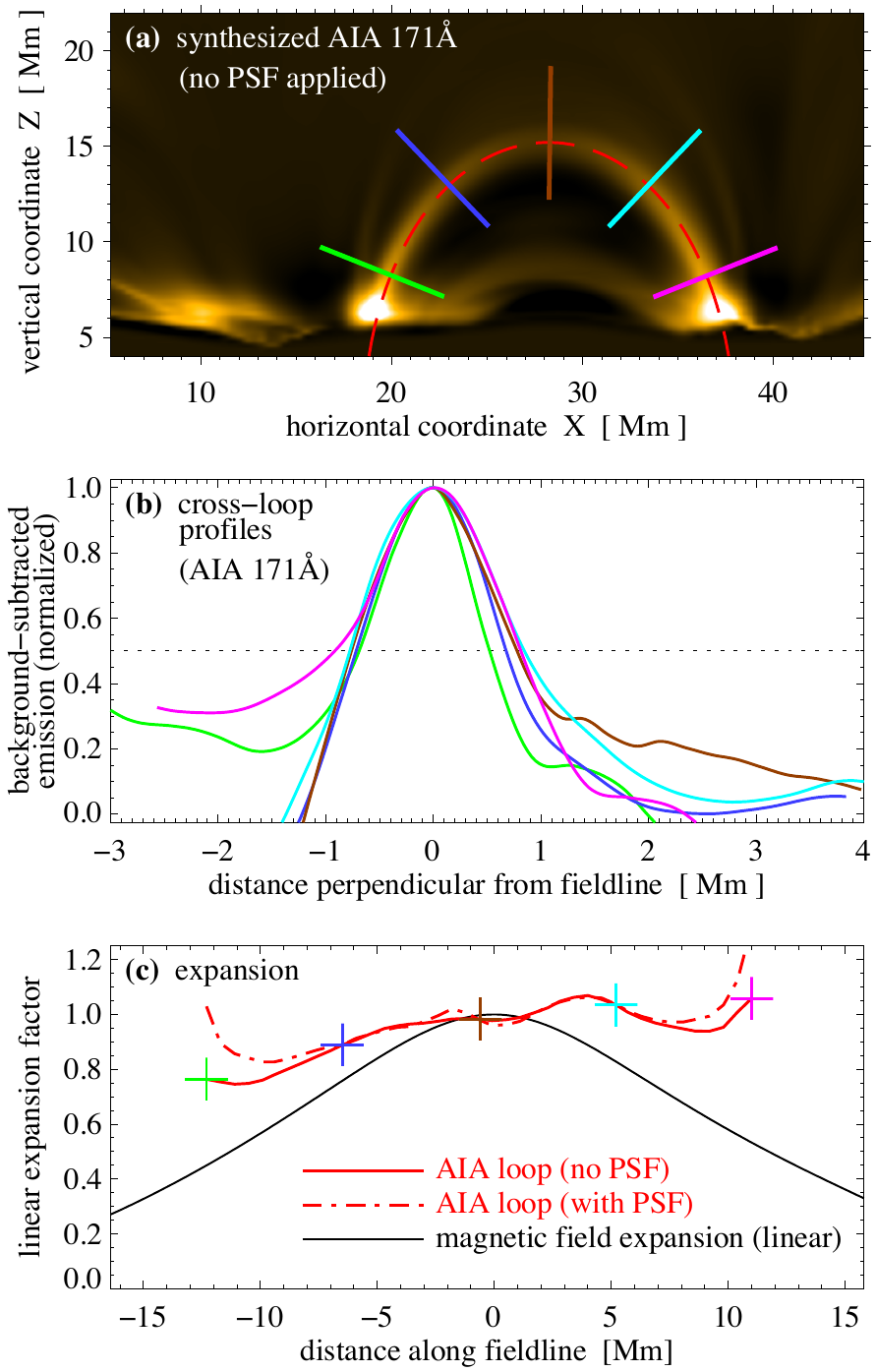}}
\caption{Analysis of the width of the synthesized coronal loop.
Panel (a) displays the emission (same as \fig{F:synt.aia}, now without the point spread function, PSF). Overplotted are the magnetic field line along the loop spine (red dashed) and selected cross sections. 
Panel (b) shows the (normalized) cross-sectional intensity profiles along the lines in panel (a) with matching color coding.
Panel (c) shows the width of the loop normalized to the width at the apex, which defines the (linear) expansion of the loop.  This expansion with (dashed) and without (solid) the PSF applied to the synthesied image is plotted in red. Overplotted is the magnetic expansion (black; ${\propto}|B|^{-1/2}$) along the spine field line. The crosses indicate the expansion at the cross-sectional profiles in panel (a).
\label{F:width}}
\end{figure}

To quantify the width of the coronal loop and its expansion we investigate the full width at half maximum (\FWHM) of the emission as a function of arc length along the loop. In \fig{F:width}a we show the same as in \fig{F:synt.aia} (now without the PSF applied, i.e.\ with the full model resolution), together with the magnetic field line along the spine of the loop. For five selected cross sections we show the (normalized) intensity profiles in panel b. The (linear) expansion of the loop is defined as the \FWHM\ normalized to the \FWHM\ at the loop top and is marked for the selected profiles in panel c by the crosses. The red solid line connecting the crosses shows the expansion all along the loop. For comparison the dashed line shows the expansion if one applies the AIA PSF. Either way, the linear expansion is small, of the order of 20\% and less. In contrast, the magnetic expansion is much stronger. Because the magnetic flux is conserved, the area expansion is proportional to the inverse of the magnetic field strength $|B|$. Thus the linear (or radial) magnetic expansion is proportional to $1/\sqrt{|B|}$, which is shown by the black solid line in \fig{F:width}c. Clearly, the (linear) magnetic expansion is well above a factor of two from the coronal base to the apex.

From this we can conclude that we see a clear expansion of the magnetic field, similar to a potential field extrapolation. This is to be expected for the flux braiding mechanism: the magnetic resistivity will prevent the build-up of too large twist through the footpoint motions and thus the magnetic field stays close to potential \citep{Galsgaard+Nordlund:1996,Nordlund+Galsgaard:1997,Schrijver:2007}. Nonetheless, the resulting synthesized EUV loop shows a constant cross section.

\begin{figure*}
\centerline{\includegraphics[width=0.9\textwidth]{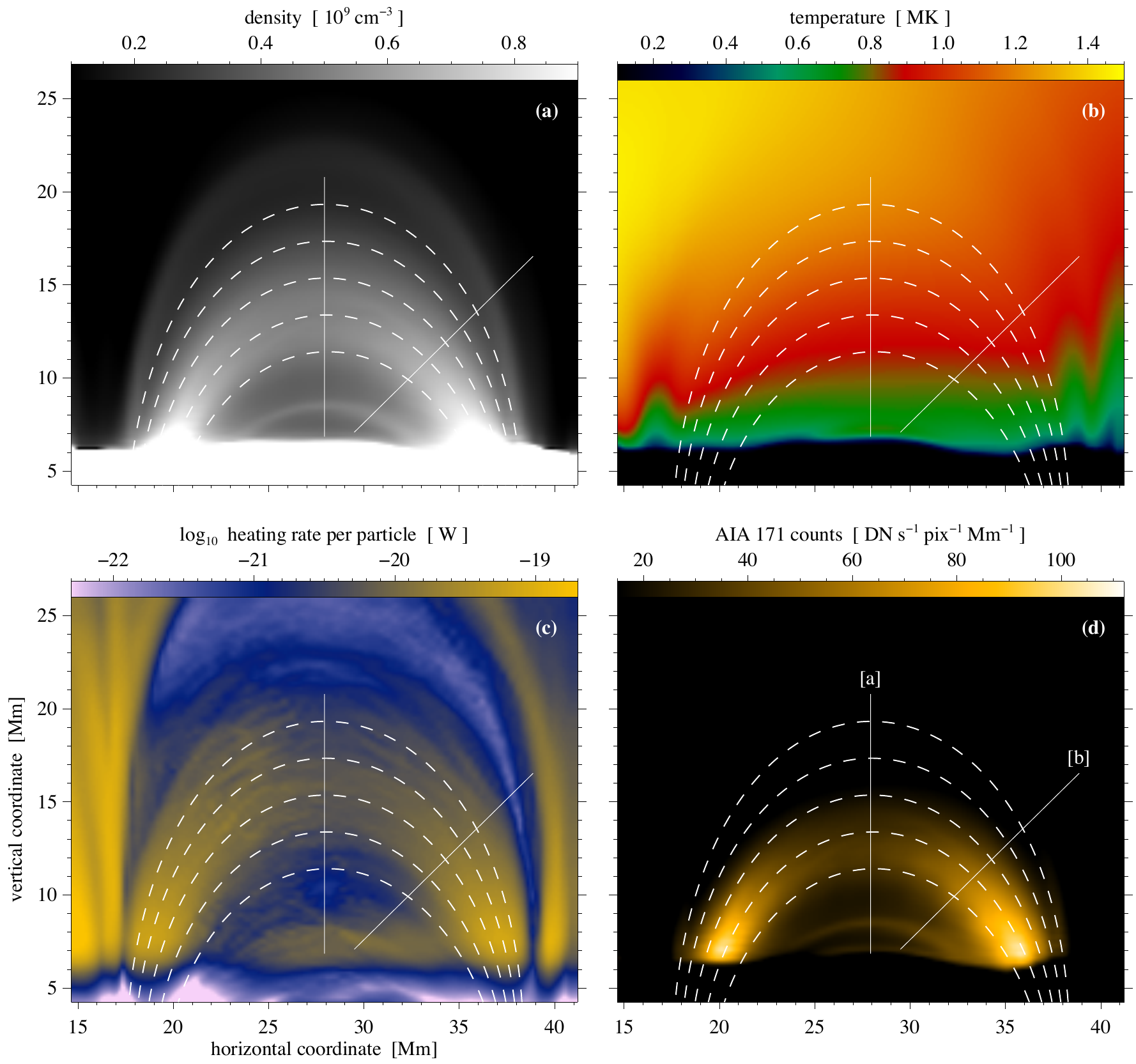}}
\caption{Two-dimensional cut through the loop plane.
Panels (a) and (b) show the density and temperaure at a snapshot at the same time as \figs{F:overview} to \ref{F:width}. Overplotted are the magnetic field lines (dashed) defining the \emph{magnetic loop}, which roughly coincides with the \emph{plasma loop} defined by the density structure.
Panel (c) displays the heating rate per particle averaged over 10 minutes before the time of the other snapshots. This also roughly coincides with the magnetic loop.
Panel (d) shows the synthesized emission in the AIA 171\,\AA\ channel defining the \emph{eLoop} outlining the EUV emission. This is smaller than the magnetic loop and the plasma loop and has roughly constant width.
The lines labeled [a] and [b] indicate the cuts shown in \fig{F:line.cuts}.
\label{F:loop.cut}}
\end{figure*}

\section{Discussion\label{S:discussion}}

To understand the nature of the constant cross section of the loop we investigate the plasma properties in a 2D cut through the loop plane which is shown in \fig{F:loop.cut}. While panels a, b, and d show the properties at a given time in the simulation, the heating rate in panel c shows the time average over ten minutes \emph{before} the snapshots in the other panels. The heating rate per particle is concentrated towards the footpoints, consistent with earlier investigations. The increased heating rate, roughly confined to the \emph{magnetic loop} defined by the field lines, leads to enhanced heat conduction back to the Sun, resulting in evaporation and filling the heated structure with mass, as visible in \fig{F:loop.cut}a. The enhanced density defines a \emph{plasma loop}, which is expanding along the magnetic field.

\begin{figure}
\centerline{\includegraphics[width=0.9\columnwidth]{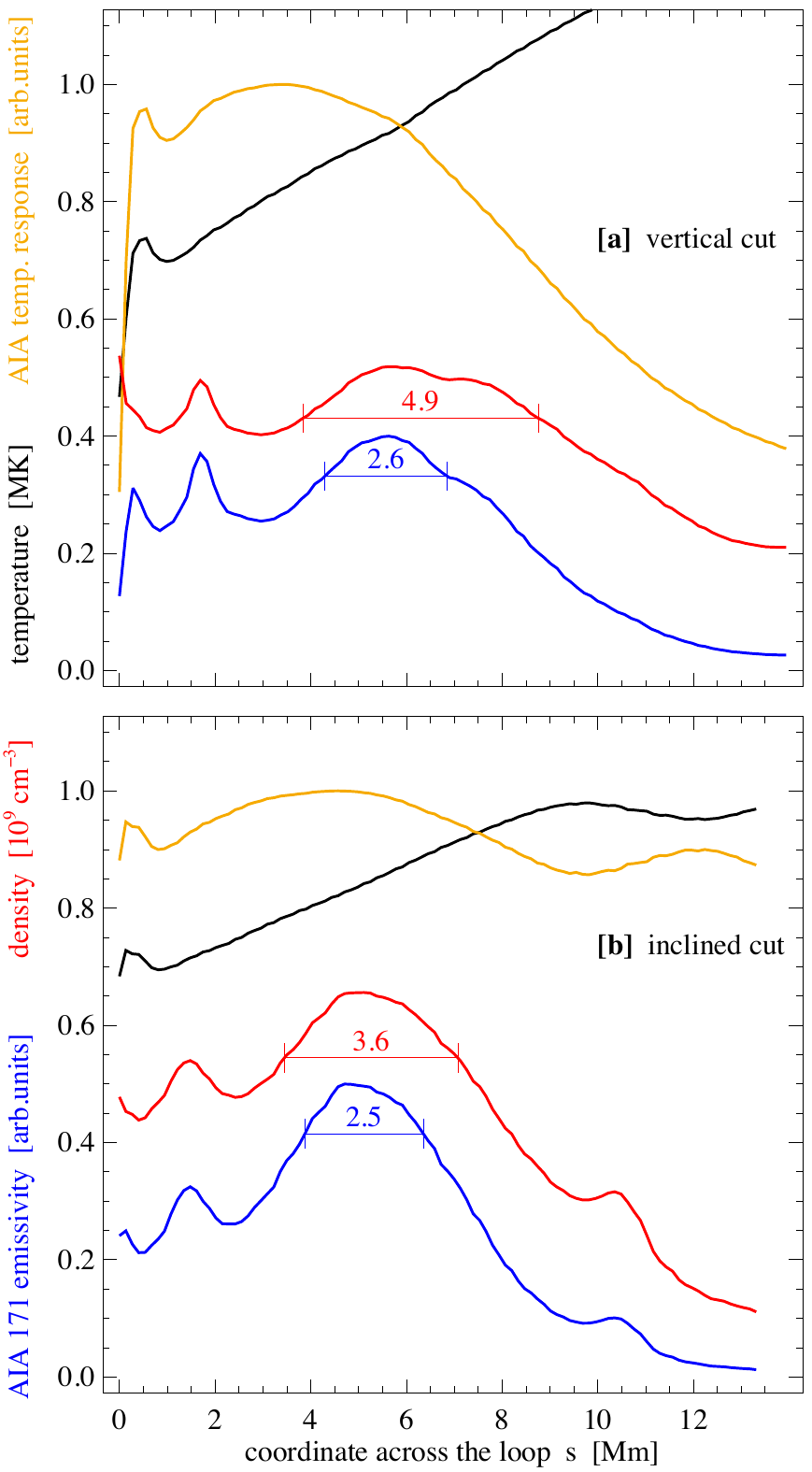}}
\caption{Profiles of AIA 171\,\AA\ temperature response, temperature, density, and AIA 171\,\AA\ emission along the cuts [a] and [b] in \fig{F:loop.cut}. The horizontal bars show the full width at 80\% of the maximum value (numbers giving the values in Mm). 
\label{F:line.cuts}}
\end{figure}

The temperature structure, however, is not aligned with the magnetic field (\fig{F:loop.cut}b). Each field line is basically thermally isolated from the neighboring field lines and along each field line the temperature rises towards the apex (similar to what is found in 1D models). Because the heating rate is not the same for each field line also the temperature on different field lines will not be the same, just as if a collection of field lines would be treated in individual 1D models. As in scaling laws \citep[e.g.][]{Rosner+al:1978} longer field lines will reach higher temperatures (for the same heating rate). This leads to a temperature gradient across the loop structure as visible in \fig{F:loop.cut}b and \ref{F:line.cuts} with the temperature increasing ``outwards''.

The AIA response function for the 171\,\AA\ channel has its maximum around 0.9\,MK, i.e.\ in the reddish region in \fig{F:loop.cut}b. In the upper part of the \emph{magnetic loop} the temperature is too hot for the AIA 171\,\AA\ channel and despite the significant density, there the AIA emission is only small. Consequently the AIA emission is confined to the lower part of the \emph{magnetic loop} (\fig{F:loop.cut}d). The resulting loop defined by the emission, the \emph{``eLoop''}, shows a more or less constant cross section.

To quantify the formation of the \emph{eLoop}, we show in \fig{F:line.cuts} cuts across the loop from \fig{F:loop.cut}. In the vertical cut [a] the temperature profile (across the field) restricts the region, which contributes to the AIA emission to the lowermost part of the density peak; consequently, the emission profile is much narrower than the density profile. For the inclined cut [b], the AIA contribution is quite flat across the density peak and thus the emission profile basically follows the density squared. Between the two cuts [a] and [b], the magnetic field strength on the loop spine drops from about 30\,G to 16.5\,G, which corresponds to a (linear) expansion of the magnetic field by some 35\%. This is basically the same for the expansion of the density structure, i.e.\ the \emph{plasma loop} (from 3.6\,Mm to 4.9\,Mm). However, the expansion of the emission defining the \emph{eLoop} is only some 3\% from about 2.5\,Mm to 2.6\,Mm. The width of the plasma loop and the eLoop has been determined by the full width at 80\%. In contrast to the analysis in \fig{F:width}, no background subtraction was applied here leading to less contrast between the region in and next to the loop.

\begin{figure}
\includegraphics{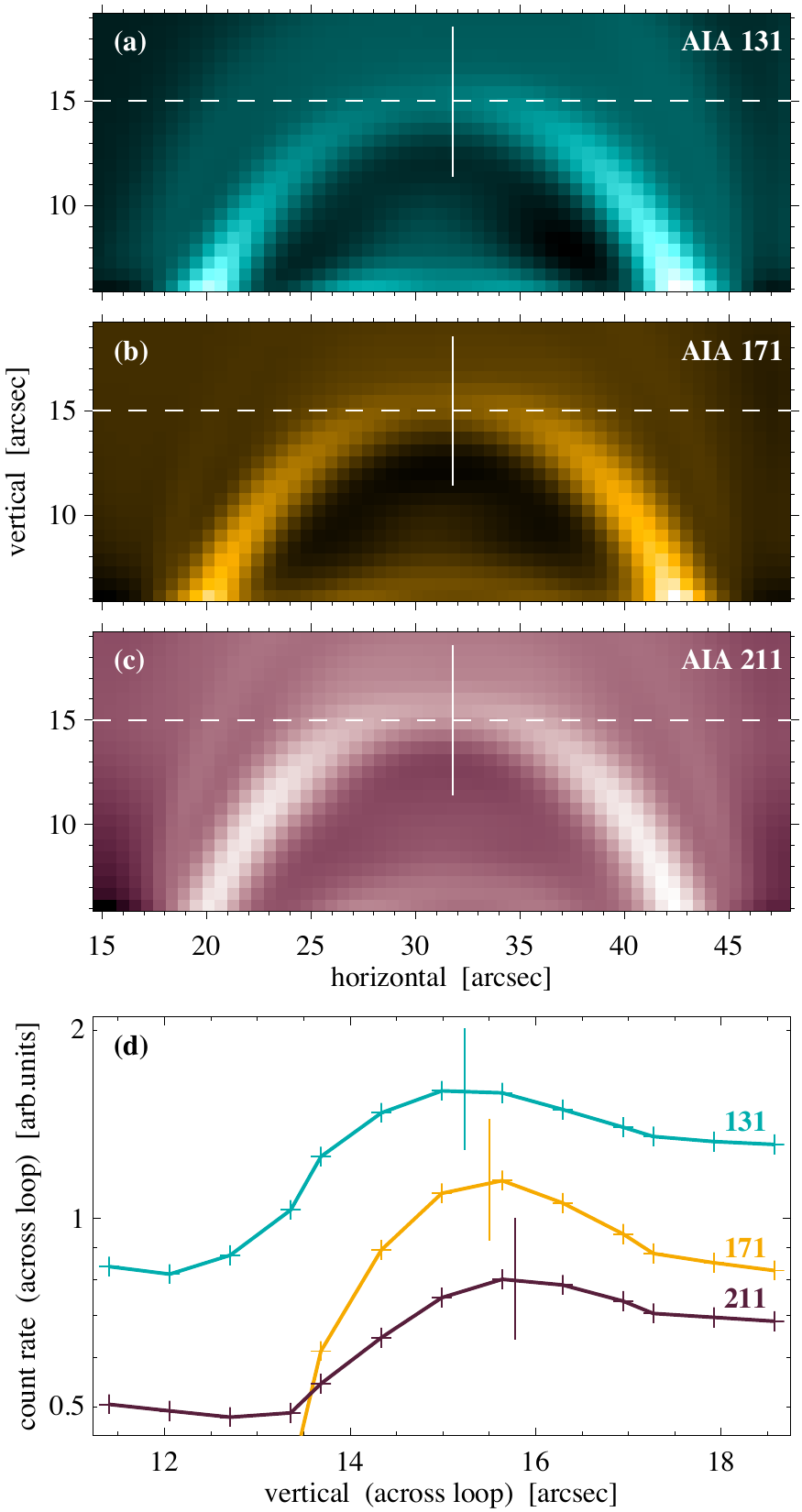}
\caption{Appearance of loop in different passbands of AIA. Panels (a) to (c) display the loop seen from the side in AIA 131, 171, and 211\,\AA. The peak contribution of these bands is at 0.6, 0.8, and 1.9\,MK, respectively. The images are shown with the AIA pixel size, and with the AIA PSF applied and background-subtracted. The horizontal dashed line is plotted for reference only. The bottom panel (d) shows the variation \emph{across} the loop near its apex as indicated by the vertical line in the top panels (ordinate plotted logarithmically). The vertical lines in panel (d) indicate the position of maximum emission in the respective band.
\label{F:loop.align}}
\end{figure}

So far we have presented results only for the AIA 171\,\AA\ channel, because this one shows the least contamination by cool material of the coronal AIA channels \citep{Martinez-Sykora:2011.aia}. If we repeat the analysis for the AIA channels that show cooler and hotter plasma, e.g. the 131\AA\ or 193\,\AA\ channels, we also see a loop at constant width (see \fig{F:loop.align}). Their cross-sectional profiles are more or less Gaussian with a width of some 2\arcsec\ to 3\arcsec, consistent with observations of, e.g., \cite{Watko+Klimchuk:2000}. They have a small spatial offset to the 171\,\AA\ loop by some 1/4\arcsec\ to 1/2\arcsec, but it is still overlapping.  The small offset in location of the loops in the passbands is consistent with observations that often show almost cospatial emission of loops in the different AIA channels \citep{Aschwanden+Boerner:2011}.

The fundamental reason for the constant cross section found in this study is the presence of a (small) temperature gradient \emph{across} the structure of enhanced density. In combination with the narrow contribution function to the EUV lines (and thus the EUV passbands of AIA), this cuts off part of the structure with enhanced density, as illustrated in \figs{F:loop.cut} and \ref{F:line.cuts}. Therefore our explanation applies not only to the coronal bandpasses of AIA discussed here, but also to the previous TRACE and EIT/SOHO data. It even applies to single line rasters with EUV spectrometers such as EIS/Hinode, because the (temperature) contribution functions of individual coronal lines typically have widths of some 0.3 to 0.4 in ${\log}T$\,[K], which is comparable to the widths of the peaks of the AIA temperature response functions \citep[cf.][]{Boerner+al:2011}. In principle this argumentation would also apply to the formation of transition region loops (seen e.g.\ in \ion{C}{4}), but the model presented here does not provide the resolution needed to properly resolve these cool structures. 

At this point we cannot say much about the mechanism that leads to the constant cross-section of loops seen in X-rays. A comparison of synthesized X-ray images to observations would be of particular interest, because the X-ray filters cover a much wider range of temperatures than the EUV filters. However, the contribution functions of the filters in current and past X-ray instruments mostly peak well above in ${\log}T$\,[K]$=$6.5 \citep[e.g.][]{Takeda:2011}. This is beyond the peak temperature we reach in the model presented here. Therefore future models reaching higher temperatures will have to show whether the explanation provided here for the EUV images will also hold for the X-ray images.

\section{Conclusions\label{S:conclusions}}

We have presented a self-consistent 3D magnetohydrodynamics model of the solar corona, which for the first time gives a detailed explanation of loops in the corona that ubiquitously show a nearly constant cross section in EUV observations. Through this we solve a problem that has remained enigmatic since the first coronal loop observations.

We find that the cross-sectional intensity profiles are close to Gaussians with widths of a few arcsec, just as found in observations \citep{Watko+Klimchuk:2000}. While the loop seen in emission has hardly any expansion, again as in observations \citep{Watko+Klimchuk:2000}, the plasma fills the magnetic structure; i.e.,\ the \emph{plasma loop} closely follows the expanding magnetic field. The constant width of the loop is then due to the variable contribution of the plasma at different temperatures. In a manner of speaking, the contribution function (depending on temperature) cuts off parts of the expanding plasma loop, and the remaining loop seen in emission then has a more or less constant cross section.
From these results it is also clear that a 1D model for a coronal loop would not be able to capture the mechanism of the formation of a constant-cross-section loop as outline here.
Furthermore, the loop width obtained through our model is consistent with EUV loop observations and a simple order-of-magnitude model for the field-line-braiding process \citep{Schrijver:2007}.

All together, this provides evidence that the process of field-line braiding with subsequent dissipation of the induced currents, which is at the heart of our model,  provides the proper spatial and temporal distribution of energy input to heat and drive the dynamic corona on observable scales. Of course, this does not exclude other mechanisms, but shows that the 3D models based on field-line braiding passed yet another observational test.

{
\acknowledgements
The authors would like to thank the anonymous referee for constructive comments. The observational data used are provided courtesy of NASA/SDO and the AIA science team.  We thank Paul Boerner for making the AIA PSF available to us before it was implemented in \emph{SolarSoft}.
}

{

}


\newpage

\begin{figure}
\includegraphics{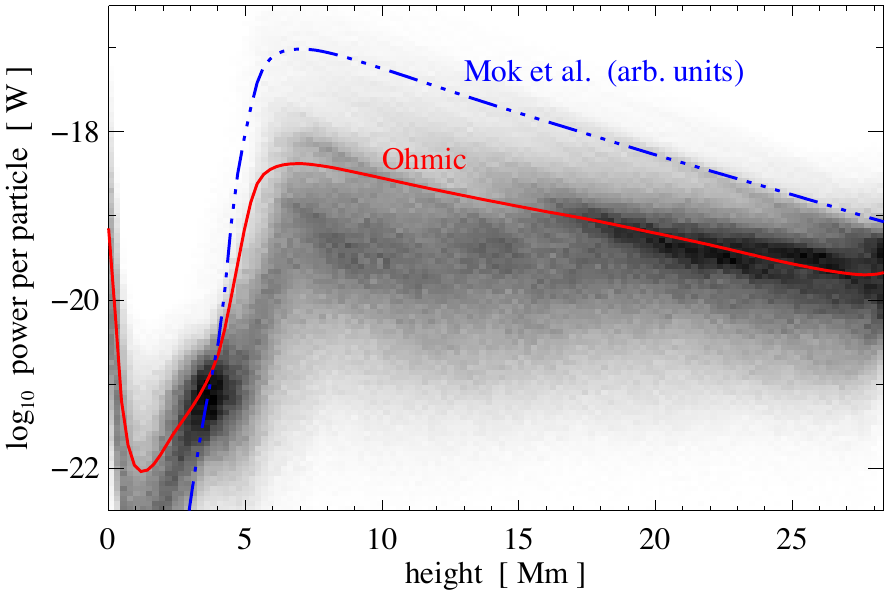}
\caption{Heating rates per particle. The 2D histogram displays the distribution of the Ohmic heating rate per particle, with the horizontal averages shown as a solid line. Overplotted as a dotted-dashed line is the horizontally averaged parameterization of the heating rate as used by \cite{Mok+al:2008}.
\label{F:heating}}
\end{figure}


\appendix

\section{Heating rate in the 3D MHD model\label{S:app}}

At the end of \sect{S:loop} we mentioned that one possible cause for the overly bright footpoint regions of the loops found in previous models \citep[][and references therein]{Mok+al:2008} might be the stronger concentration of the heating rate towards the footpoints in the \cite{Mok+al:2008} model. To substantiate this, we give a short comparison of the heating rates in our model and in \cite{Mok+al:2008}.

In our model the energy input is due to Ohmic heating through the dissipation of currents $j$ induced by field line braiding with a constant magnetic resistivity $\eta$ at a rate 
$$
H_{\rm{Ohm}}=\eta\,\mu_0~j^2~.
$$
In \fig{F:heating} we show the 2D histogram of the Ohmic heating rate \emph{per particle} (volumetric heating rate divided by the particle density)
in our model as a function of height in the atmosphere. For this we evaluated $H_{\rm{Ohm}}$ at each grid point for one snapshot. The picture does not change (significantly) over time. Overplotted as a thick solid line is the horizontally averaged Ohmic heating per particle. As emphasized in previous studies, the heating per particle shows a peak at the base of the corona.

In their model \cite{Mok+al:2008} use a parameterization of the heating rate through the local magnetic field strength $B$ and density $\rho$,
$$
H_{\rm{Mok}} \propto B^{2.5} ~\rho^{-0.5}~.
$$
Again, we evaluate $H_{\rm{Mok}}$ at each grid point and overplot the horizontal averages of  $H_{\rm{Mok}}$ per particle as a function of height in \fig{F:heating}.

It is clear from \fig{F:heating} that the heating rate per particle as used by \cite{Mok+al:2008} shows a stronger concentration towards the base of the corona. This could be part of an explanation of the too strong emission near the loop footpoints as found by \cite{Mok+al:2008}.

\end{document}